\RequirePackage{etoolbox}
\csdef{input@path}{{style/}{graphics/}}
\documentclass[MSNbibl,nameyear,dvips]{arxstspdf}
\usepackage{graphicx}

\usepackage{flushend}
\usepackage{stfloats}

\volume{30}
\issue{1}
\pubyear{2015}
\firstpage{118}
\lastpage{132}
\doi{10.1214/14-STS492} 
\docsubty{FLA}


\begin{document}
\begin{frontmatter}

\title{A Conversation with Richard A. Olshen}
\runtitle{A Conversation with Richard A. Olshen}

\begin{aug}
\author[A]{\fnms{John A.}~\snm{Rice}\corref{}\ead[label=e1]{rice@stat.berkeley.edu}}
\runauthor{J. A. Rice}

\affiliation{University of California, Berkeley}

\address[A]{John A. Rice is Emeritus Professor, Department of
Statistics, 367 Evans Hall,
University of California, Berkeley, California 94720-3860, USA \printead{e1}.}
\end{aug}

\setattribute{abstract}{width}{35pc}

%
\begin{abstract}
Richard Olshen was born in Portland, Oregon, on May 17, 1942.
Richard spent his early years in Chevy Chase, Maryland, but has lived
most of his life in California. He received an A.B. in Statistics at
the University of California, Berkeley, in 1963, and a Ph.D. in
Statistics
from Yale University in 1966, writing his dissertation under the
direction
of Jimmie Savage and Frank Anscombe. He served as Research Staff
Statistician and Lecturer at Yale in 1966--1967.

Richard accepted a faculty appointment at Stanford University in 1967,
and has held tenured faculty positions at the University of Michigan
(1972--1975), the University of California, San Diego (1975--1989), and
Stanford University (since 1989). At Stanford, he is Professor of
Health Research and Policy (Biostatistics), Chief of the Division of
Biostatistics (since 1998) and Professor (by courtesy) of Electrical
Engineering and of Statistics. At various times, he has had visiting
faculty positions at Columbia, Harvard, MIT, Stanford and the Hebrew University.

Richard's research interests are in statistics and mathematics and
their applications to medicine and biology. Much of his work has
concerned binary tree-structured algorithms for classification,
regression, survival analysis and clustering. Those for classification
and survival analysis have been used with success in computer-aided
diagnosis and prognosis, especially in cardiology, oncology and
toxicology. He coauthored the 1984 book \textit{Classification and
Regression Trees} (with Leo Brieman, Jerome Friedman and Charles Stone)
which gives motivation, algorithms, various examples and mathematical
theory for what have come to be known as CART algorithms. The
approaches to tree-structured clustering have been applied to problems
in digital radiography (with Stanford EE Professor Robert Gray) and to
HIV genetics, the latter work including studies on single nucleotide
polymorphisms, which has helped to shed light on the presence of
hypertension in certain subpopulations of women.

Richard also has a long-standing interest in the analyses of
longitudinal data. This includes a detailed study of the
pharmacokinetics of intracavitary chemotherapy with systemic rescue
(with Stephen Howell and John Rice). Related efforts have focused on
``mature walking,'' concomitants of high cholesterol, and aspects of
glomerular filtration in patients with nephrotic disorders (with Bryan
Myers). With the late David Sutherland, Edmund Biden and Marilynn
Wyatt, he coauthored the monograph \textit{The Development of Mature
Walking}. Richard's other stochastic-statistical interests include
exchangeability, conditional significance levels of particular test
statistics, CART-like estimators in regression and successive
standardization of rectangular arrays of numbers.

Richard is an elected Fellow of the IMS, the AAAS, the ASA and the
IEEE. He is a former Guggenheim Fellow and has been a Research Scholar
in Cancer of the American Cancer Society.
\end{abstract}

\end{frontmatter}

\textbf{Rice:} It's a great pleasure to interview you, Richard. We go
back a long way to UC San Diego in the 1970's. I'd like to begin with
your distant past, if your memory goes back that far. What childhood
influences drew you into mathematics and statistics and medical science?

\begin{figure}

\includegraphics{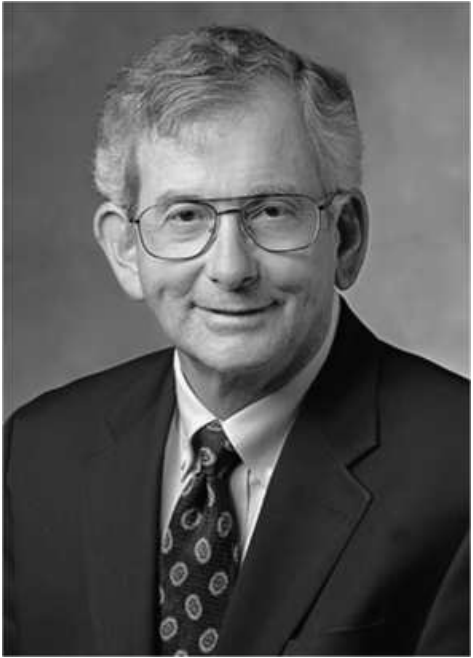}

\caption{Richard Olshen, 2010.}\label{fig1}
\end{figure}

\textbf{Olshen:} My father was a Ph.D. student of Henry Lewis Rietz of
the Rietz Lectures and the IMS. He had a Ph.D. in mathematics from the
State University of Iowa. He was a very troubled person, but every once
in a while he had a clear view of things. He knew a fair bit of
mathematics and some statistics, such as it was when he was younger.
Interesting ideas were always in the air. I remember when I was a child
wondering if there were more real numbers than integers.

\textbf{Rice:} At what age was that? Do you know, roughly?

\textbf{Olshen:} I think I was nine. I remember learning something
about transfinite arithmetic and Cantor; but that was many years ago,
and I don't remember many of the details. As far as statistics goes, I
was a joint Statistics and Mathematics major at Berkeley, but the
person who advised my letter of the alphabet in the Department of
Mathematics wanted me to take a course that I didn't want to take; so I
dropped the Mathematics part.

\textbf{Rice:} Oh, is that how you ended up a Statistics major? I know
you went to Berkeley, presumably because you had an interest in Mathematics.

\textbf{Olshen:} I was recruited to go to the University of Chicago for
no good reason I could ever discern. My father wouldn't hear of me
going to the University of Chicago. A woman of whom I was very fond was
going to Berkeley, and I thought it was a pretty good school. It wasn't
Stanford. Stanford was not a welcome word in my house.

\textbf{Rice:} Oh, really?

\textbf{Olshen:} When I was a junior in high school, my mother and I
went to college admissions night at Burlingame High School. We lived in
Burlingame, California, then, which is near the San Francisco airport.
We were sitting in a room, and the woman who was somehow in charge of
outreach from Stanford got up; and the first words out of her mouth
were, ``Life doesn't end if you don't get into Stanford.'' My mother
grabbed my arm and pulled me out of the room and said, ``You're not
applying there.''

\textbf{Rice:} Good for your mother!

\begin{figure}[b]

\includegraphics{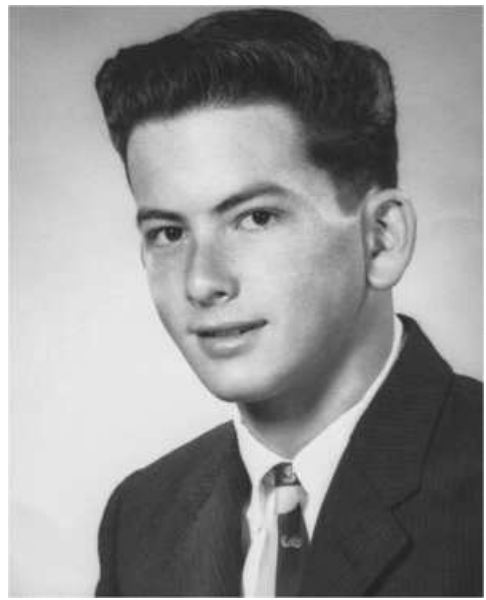}

\caption{Richard's high school graduation picture. Taken Spring 1958,
age 16.}\label{fig2}
\end{figure}

\begin{figure}

\includegraphics{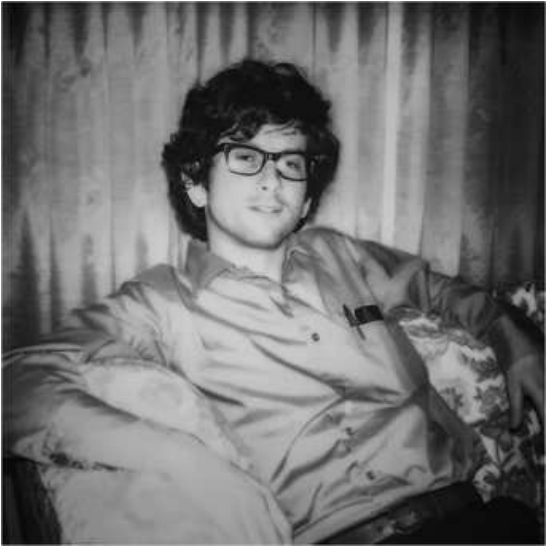}

\caption{Richard as a graduate student at Yale. Taken Spring 1965.}\label{fig3}
\end{figure}

\textbf{Olshen:} That didn't appeal to her aesthetic.


\textbf{Rice:} What was the course in statistics that drew you into the
subject at Berkeley?

\textbf{Olshen:} It wasn't so much a course as it was a person. I had
taken probability when I was a sophomore. In those days, you could
actually do Volume 1 of Feller. Now it's somewhat forbidden because
the problems are too hard. I was pretty good at it. Then, when I~was a
junior in college I met the late David Freedman. I believe I was in the
first course he taught at Berkeley.

\textbf{Rice:} That must have been very early in his career.

\textbf{Olshen:} That was in 1961. He was somebody I~wanted to please.
He was a stern guy and obviously very sharp. In those days, it was
clear, both at Berkeley and at Yale, that the very young faculty---and
David was certainly young then---that the young faculty looked on the
able students as competitors for their jobs, and so that tension was
always there. He was pretty secure and didn't really feel this way, but
there were others who seemed to have that attitude.

\textbf{Rice:} You thought that at both Berkeley and Yale?

\textbf{Olshen:} Well, those were two good places to get jobs. Yes, I
was surprised, but that was the sense that I~had, especially at Yale.

\textbf{Rice:} How many statistics majors were in your class? It must
have been a very small number.

\textbf{Olshen:} I don't know. There were more than 10 and not more
than 30.

\textbf{Rice:} It's grown substantially in the last few years. There
are more than 300 now at Berkeley.

\textbf{Olshen:} There's a change in enrollment at Stanford, too,
although there is no undergraduate major in statistics at Stanford.
There's a Math/Comp Sci major on which Bradley Efron has worked hard.
It's really good, and it's one of the best undergraduate majors at
Stanford. I don't know how many students it has, but quite a few. The
master's program in the Department of Statistics was almost nonexistent
10 years ago. Now it has 90+ students, and people are clamoring to get
into the master's program. People from all over the world who would
have been Ph.D. students 20 years ago, bright kids.

I think statistics serves these young people well. It teaches them
something about computing. It teaches them something about statistical
inference. I think these are all good things to know, no matter what
they choose to do. It's hard to learn much subject matter very well if
you're less than 20, but in your early twenties it's a good idea to
learn what you can. That means juniors and seniors in college and first
and second year graduate students.

\textbf{Rice:} That's what you did in going from Berkeley to Yale. Was
that transition a big change?

\textbf{Olshen:} Well, yes and no. It was not a change in difficulty.
They were equally difficult. Berkeley and Yale were both, for me,
really, really hard. If they had been one percent harder I couldn't
have done either one. But Berkeley's statistics was more decision
theoretic. I would say more of Wald's descendents than anything else.
Yale, because Frank Anscombe was the founder of its Department of
Statistics, was much more British, much more Fisherian, much more
likelihood oriented. I was the first Ph.D. student there.

What was deemed important in statistics was different in the two
places. But as a student, one is faced with challenges of various
sorts, and those challenges were formidable for me in both places.

\textbf{Rice:} What led you to go to Yale for your graduate studies?

\textbf{Olshen:} Well, I thought, perhaps, that I would go to
Princeton; and David Freedman, who influenced me at Berkeley, was a
friend of Frank Anscombe, who was then at Princeton. Anyway, I visited
Princeton the summer of 1962, and ultimately was not admitted to the
Department of Mathematics at Princeton, anyway. But I didn't know that
then, when in January of 1963, I got a personal letter from Frank,
which included an application to Yale. Frank said, ``I'm moving from
Princeton to Yale. Do you want to come to Yale? If you do, here's an
application.'' I asked David, ``Is there anybody good at Yale?''
because I didn't know much about it. For personal reasons, I wanted to
get far away from the San Francisco Bay area. I thought New Haven was
far enough.

David said, ``Oh, yeah. There are lots of great people at Yale.'' He
mentioned some of them. He was right. I~said, ``Fine. I'll go there.''
That's how I ended up there.

\textbf{Rice:} Which Yale faculty had a strong influence on you?

\textbf{Olshen:} During my four years in New Haven, Yale in general and
Hillhouse Avenue there in particular were exciting places to be. Frank
Anscombe was a remarkable statistician who, in retrospect, kept his
considerable mathematical skills too hidden. Several of us, especially
Frank and Phyllis Anscombe, recruited Jimmie and Jean Savage to Yale in
the spring in 1964. Jimmie's last name, not his original family name,
was nonetheless well chosen. I believe that his fame in statistical
history is deserved. Especially my last two years in New Haven he was
extraordinarily generous with his time, usually spending an hour with
me every day, most time spent working on mathematical problems.
Jimmie's abilities were remarkable. For example, unaware of the work of
Kolmogorov and Arnold before him, Jimmie solved a variation of
Hilbert's thirteenth problem by himself. Unfortunately, few solutions
of the various problems we discussed ever led to publications. The late
Shizuo Kakutani taught measure theory and many aspects of probability.
 Paul L\'{e}vy was the originator of much probability
Kakutani taught; so, too, were the studies of Markov processes and
ergodic theory by him and Yoshida. Some of the problems in measure
theory he asked us owed to (separate) books by Kuratowski, Sierpinski,
and Hausdorff, though students were left to discover them ourselves.
Alan James taught multivariate analysis from his unique perspective
that combined as serious computation as was possible then with the
study of matrix groups. There was a year long course in utility theory
and game theory by Johnny Aumann, and much else, too.

\textbf{Rice:} Let me look ahead in time here. Your career has had a
remarkable trajectory. I think one of your first publications was on
asymptotic properties of the periodogram.

\textbf{Olshen:} That was my thesis.

\textbf{Rice:} Oh, I didn't know that. One of the most recent was on
some cytokine bead assays. I don't know even what they are. But I
wonder, before we go into some of these areas in more detail, as an
overview, are there landmark topics that you've visited during your
career that sketch out the contours of this trajectory? We'll return to
some of these more in depth later, I'm just trying to get a sense of
scope and flow now.

\textbf{Olshen:} I think of statistics as a triangle. There's a
computational part, at which you're very expert, and I'm not; a
mathematical part, which I think is one of my strengths; and subject
matter stuff. I can't do all three corners of the triangle, or at least
I'm not very good at one of them; and so I try to do the other two. I
grew up in the Sputnik era. Mathematics was one of those things that
was in the air, and so that's what we did. When I was a freshman in
college, I remember being in a class where we did Hardy's \textit{A
Course in Pure Mathematics}. We tried to do the problems, which were
pretty tough for this 17-year old.

\textbf{Rice:} Yes, that was an era of intense interest and enthusiasm
for mathematics and mathematics education. It was a heady time to be a
math major.

\textbf{Olshen:} As far as subject matter goes, my feeling is that it's
hard for me, maybe because I'm slow, to be much of a dabbler. I've
encountered a few topics that have really interested me, and I've tried
to stay with them long enough so that I could learn enough to be of
use. I think that if you're going to do statistics, then you have to
meet subject matter people on their turf. In order to do that, you have
to eat humble pie, a lot of it sometimes, and be willing to take your
lumps, and just try your best to learn whatever subject it happens to
be. There have been four or five subjects in my life that I've tried to
learn. Probably I've not learned any of them very well, but it's not
for lack of trying.

That's always been my attitude. There was the mathematics on the one
hand, and there was trying to learn subject matter areas on the other.
Together, they've been pretty much a full time job.

\textbf{Rice:} After you got your Ph.D., you moved around a bit, spending
time at Columbia, Michigan and Stanford. Then you landed in San Diego
in 1975. What led you to come to San Diego? I was very happy you did,
of course.

\textbf{Olshen:} Well, I was happy, too. There were a couple reasons.
First of all, they would have me, which was not a trivial matter.
Second of all, they gave me tenure  {ab initio}. Since I had had
tenure at the University of Michigan anyway, and offers of tenure at
other places, that was important to me. When I came to San Diego my
billet, or whatever it was called, was joint between Mathematics and
the School of Medicine. I was interviewed by the Dean of the Medical
School, who asked, ``Are you really interested in medicine?''

I was interested enough to say, ``If you hire me, I'll be faithful to
the medical school's welfare.'' I meant it, and I tried to be. The idea
of doing mathematics and medicine always appealed to me.

\textbf{Rice:} Having a foot in each of these places on campus didn't
create a cognitive dissonance?

\textbf{Olshen:} I don't know. In that respect, nothing has changed
very much. My job titles have changed, but nothing about me in that
respect has changed. I never stopped to ask. I think that people are
driven to do what they're going to do. It's not fruitful to ask why.
One does what one does. If it's robbing banks or hurting people, that's
not an admissible strategy; but you can look after your career and
pursue what interests you or do what you think you can do. I've never
stopped to ask.

\textbf{Rice:} One thing you did at UCSD during the time you were there
was to create a real presence for statistics, particularly in the
medical school, in which it hadn't had much presence before. I was
wondering: how did you go about doing that? It can be socially and
culturally difficult.

\textbf{Olshen:} UCSD was started, as you probably know, as a
university campus in the 1960s, as opposed to being merely the Scripps
Institute of Oceanography, which had existed since the early 20th
century. It was founded by Roger Revelle, an amazing man. He fought
hard against prejudices that were ruled illegal by the 1964 civil
rights law. He brought scientific activity to a part of the world where
it hadn't been so much before.

But if Roger Revelle had a blind spot, it was that he just didn't like
statistics. There was never a Department of Statistics at UCSD like
there were at various other UC campuses, as you well know.

A lot of problems in medicine really involve statistical issues, and
not just in medicine, but in a lot of scientific areas. Think of the validation of the
discovery of the Higgs boson, for example. It seemed to me that there
was a vacuum, that there was a need for people interested in
interpreting data. I don't know that I filled it very well.

\textbf{Rice:} There must have been a few key people in medicine who
helped you fill that vacuum.

\textbf{Olshen:} There were. One of the things that helped promote that
was that in the late 1970s there was an attempt to get National Cancer
Institute designation for a Cancer Center at UCSD. The leader of the
effort was John Mendelsohn. There was a group of people including not
only Mendelsohn, but also Steve Howell, Mark Green and Ivor Royston.
They were eclectic, but real dynamos, all of them in their own ways.

They included me. I think that was certainly one path. Another path
that I think was really helpful to me at UCSD was that UCSD had this
tradition of cardiovascular medicine. Gene Braunwald of Harvard had
been at UCSD briefly. He brought John Ross and Jim Covell and other
people there. There was this huge presence in cardiology. John Ross was
the leader of it when I was there. Many of these people were really
smart. They operated on dogs and what have you, so it was a little
grisly what they did. I felt I'd learned from them. It was a pleasure
to be involved in their projects. There was something called the
Specialized Center for Research in Ischemic Heart Disease, and they
included~me.

A third avenue was the Gait Lab in Children's Hospital and Health
Center. Again, that was interdisciplinary. It involved a surgeon, an
engineer and a nurse; I was the fourth of them. We didn't publish many
things, but I think what we did was pretty good.

\textbf{Rice:} Yes, your work on gait was an important early stimulus
to the development of functional data analysis.

\textbf{Olshen:} Those were three areas that I think were enabling to
me. There were many other good things at UCSD that came later.
Psychiatry is a big deal at UCSD, and eventually I got involved in the
Center for Neurobehavioral AIDS. Anyway, those were some of the
avenues. The thing they all had in common is that I had much to learn.

\textbf{Rice:} In the Department of Mathematics, where you had your
other foot, what people did you learn from especially?

\textbf{Olshen:} Well, of course, coming to UCSD, I was grateful
because Ingram Olkin at Stanford had spoken with Murray Rosenblatt.
Ingram didn't give me any reason to be optimistic, but Rosenblatt was
the senior person in the statistical community at USCD, and the whole
reason I got interested in periodograms in the first place owed to the
famous book by Grenander and Rosenblatt.

\textbf{Rice:} I remember that you knew that book quite well.

\textbf{Olshen:} Well, I had read it from the first letter to the last.
I can't say that I memorized it, but pretty close. Murray was there. He
was certainly an influence. I~knew that Adriano Garsia was at UCSD. He
had given basically a two line proof of the maximal ergodic theorem; it
led to a quick proof of the ergodic theorem, which is something that
had begun at Yale in some sense with Josiah Willard Gibbs. I had a
Josiah Willard Gibbs Fellowship at Yale when I came there, so I felt
some connection with that work. Michael Sharpe was somebody I had known
since graduate school.

\textbf{Rice:} Oh, that's right. He was a graduate student at Yale,
too, wasn't he?

\textbf{Olshen:} He was the first person I met in New Haven. I remember
talking to Michael, who was from Tasmania, which seemed like it was
pretty far away. He had been an honor student. I guess in their system,
you did three years of college, and then if you were really good, you
did a year of honors; he had done honors with the celebrated E. J. G.
Pitman, father of your celebrated colleague Jim Pitman. I remember
coming home after spending about a half hour in the Yale Co-op chatting
with Michael, and I remember telling Vivian, my wife at the time, ``If
everybody around here is as good as this guy, I'm in big trouble.''

Michael was very well educated, and he was quite smart, and that was
evident, I would say, after about 45 seconds. After 30 minutes, I was
thoroughly intimidated. I remember that Michael detested the cold in
New Haven; he came to San Diego in part because he read through books
on temperatures in the continental United States, and he wanted a high
average temperature and as small a difference as possible between the
max over the month and the min.

\textbf{Rice:} San Diego is pretty much an optimum in that metric in
the US.

\textbf{Olshen:} He said, ``I'm going there,'' and he did. Anyway, and
of course, you were there, and you were interested in time series and
all that stuff. I didn't feel like Stanford was the right place for me
to be pursuing that. There were a lot of reasons why UCSD seemed like a
good place. There were a lot of very bright, very able people.

However, I think there was a downside in that San Diego got to be a
really good place because it rapidly hired a bunch of people who were
very good, but who were unhappy where they were. They weren't unhappy
where they were because of where they were; they were unhappy with the
place because of who they were.

The medical school actually was different from some of the rest of the
campus, because as medical schools go, the medical school wasn't very
cranky. Or at least I~didn't perceive it as being so.


\textbf{Rice:} One of the best things that happened to you at San Diego
was that you met and married Susan and expanded your family.

\textbf{Olshen:} Yes, well, I was in a pretty sorry shape. I~was a
single parent.

\textbf{Rice:} How did you meet?

\begin{figure}

\includegraphics{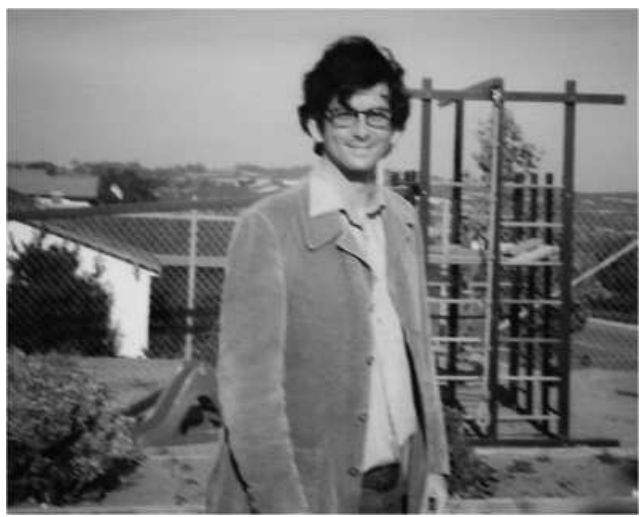}

\caption{Richard, taken in the backyard of his home in Del Mar, CA, in 1977.}\label{fig4}
\end{figure}

\begin{figure}[b]

\includegraphics{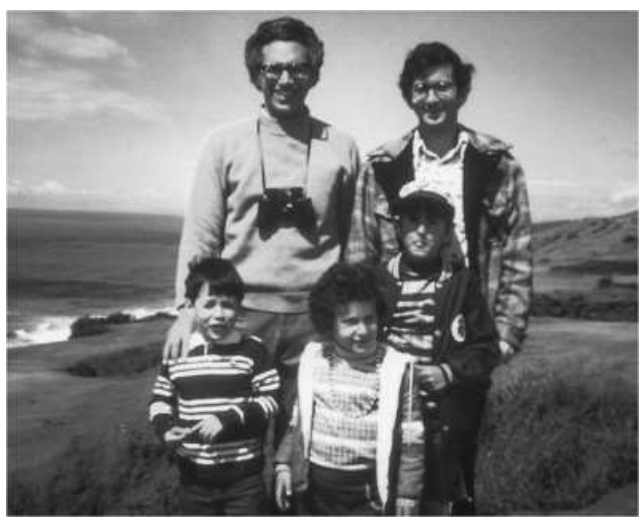}

\caption{From left to right: David Perlman, Michael Perlman, Elyse
Olshen, Adam Olshen, Richard Olshen. Picture taken in 1978 in La Jolla, CA.}\label{fig5}
\end{figure}

\begin{figure}

\includegraphics{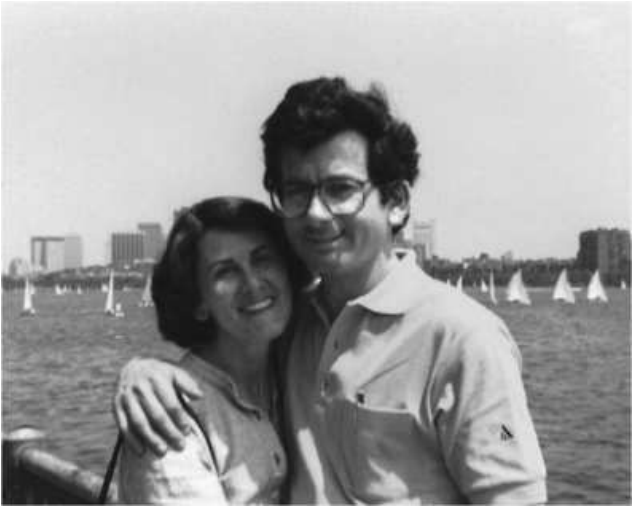}

\caption{Richard and Susan Olshen on the banks of the Charles River,
Spring 1980.}\label{fig6}
\end{figure}

\begin{figure*}[b]

\includegraphics{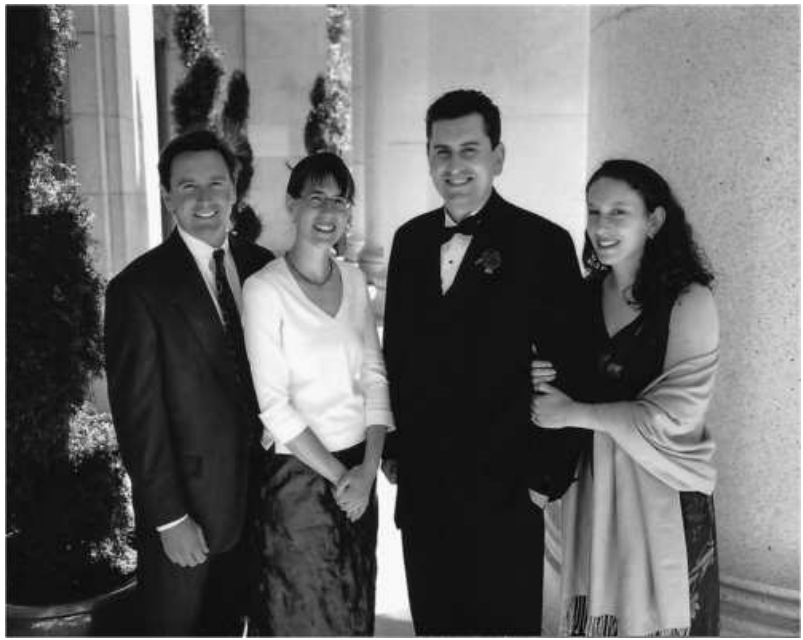}

\caption{From left to right: step-son Stephen Heller, step-daughter
Rachel Miller, son Adam Olshen, and daughter Elyse Olshen Kharbanda,
taken at Adam's wedding to Manisha Desai in 2001.}\label{fig7}
\end{figure*}

\textbf{Olshen:} Oh, I met Susan because I was a single parent living
in Del Mar Heights. There were two women in the neighborhood, Sandy
Peterson and Gail Goldberg. They used to help me, because I didn't know
about the Hebrew school, I didn't know about piano lessons; I didn't
know about soccer teams. If something came up, I would ask one of them,
``Should my child go to this school or that school or this team or this
teacher or whatever?'' One day Gail said to me, ``Richard, my husband's
partner's wife has a friend, Sue Heller, in La Jolla; and she's
separated from her husband; and if you don't call and ask her out for
dinner, I'll never speak to you again.'' So I called her.

\textbf{Rice:} That's a forceful matchmaker!

\textbf{Olshen:} I said, ``Sue Heller is the name of the wife of my
pediatrician.'' I said to Susan, ``If you're the wife or former wife of
my pediatrician, then I'm not going near you with a ten-foot pole
because one of the few things that's going well in my life is the
pediatrician. I really like this guy. He takes good care of my
children, and I like him. So if you're that Sue Heller I don't want to
get anywhere near you.'' She said enough to preclude her being the wife
of the pediatrician. I said, ``Well, OK. Do you want to go to Lescargot
for dinner?'' She was taken aback because it was a nice restaurant. But
the thing about it was this: it was really awkward for me to go there
by myself. I will say Susan was totally flabbergasted when I took her
there but I said, ``It has nothing to do with you, I like this place
and I can't come here by myself.'' Gallant I was.

So I met Susan in 1977. We got married in 1979. What's this, 2013? It's
been awhile.


\textbf{Rice:} It certainly has. In 1977, changing the topic a bit, you
were beginning to get involved with CART, which at that time\ldots

\textbf{Olshen:} Oh it was before then.

\textbf{Rice:} At that time, it seemed to me quite novel and esoteric.
Now it's a very standard tool that everybody learns; it's widely used.

\textbf{Olshen:} I started in with CART in 1974, at the Stanford Linear
Accelerator Center. I was in the Computation Research Group. Jerry
Friedman was my boss there, and he was very interested in binary tree
structured rules. They started out as rules for quick searches because
you can imagine if you want to find a nearest neighbor and you build a
tree down to where there's one observation per terminal node, you're
going to be able to find nearest neighbors pretty easily. Jerry was
interested in using this for classification. I got interested in the
application side, which had to do with a lead and plastic sandwich of
particles originating in a bubble chamber. Also, Lou Gordon and I
worked on the mathematical side.

By 1977, I was into CART. There was no book then. The book didn't come
until six or seven years later, depending on how you count.

\textbf{Rice:} In 1977, weren't Leo Breiman and Chuck Stone also involved?

\textbf{Olshen:} Leo and Chuck were definitely involved. There were
basically three groups of two, Jerry and Larry Rafsky, Chuck and Leo
and Lou Gordon and I. Larry Rafsky was busy with other things and
didn't really pursue this very extensively. Lou somehow never became
part of the milieu, but I became friendly with Chuck because I had
known him in my probability life. David Siegmund and I had worked on a
problem that Chuck ended up doing. Then in 1975 there was a meeting at
UCLA where nearest neighbors and trees and what now are called support
vector machines, but in those days were called variable width kernels,
were very much in the air. There was in CART history a famous technical
report that came in 1979 from a place where Leo did consulting in Santa
Monica and where he dragged Chuck. It was called Technology Services
Corporation.

\textbf{Rice:} I remember seeing that report. It was quite something,
very forward looking, for its time.

\textbf{Olshen:} Chuck was pretty well versed in trees several years
before then. I remember he had written a paper that he submitted to
\textit{The Annals of Statistics}. Richard Savage was the editor. He
had showed it to John Hartigan who didn't speak well of it, I guess. I~called up Savage and gave him a piece of my mind, not that I had any to
spare, and not what he wanted to hear.

\begin{figure*}[b]

\includegraphics{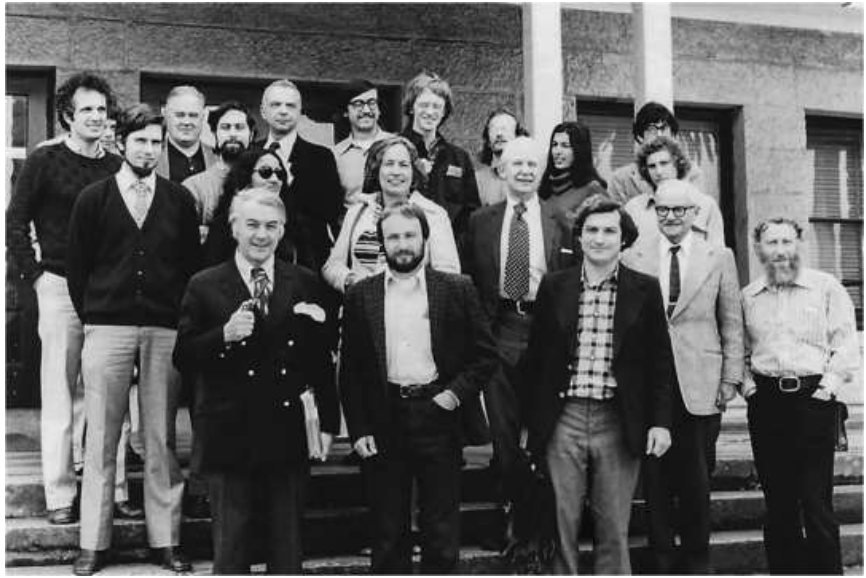}

\caption{This photo was taken in front of the old Sequoia Hall at
Stanford in 1975. The occasion was a gathering to discuss what role
statistics research might have in environmental problems. The cast of
characters is:
Back row (left to right): Brad Efron, John Tukey, Paul Switzer, Herb
Robbins, Tom Sager, not identified, Ray Faith, not identified, Richard
Olshen.
Middle row (left to right): Don McNeill, Yash Mittal, Elizabeth Scott,
Don Thomsen, Gary Simon.
Front row (left to right): Geoff Watson, Peter Bloomfield, Persi
Diaconis, Jerzy Neyman, Ingram Olkin.}\label{fig8}
\end{figure*}

\textbf{Rice:} Weren't you a discussant of that paper?

\textbf{Olshen:} Yes. It went from being rejected to being a discussion paper.

There were trees involved in that, and there were some personal
rivalries that were buried in the discussion, Lou Gordon's and my first
paper on CART for classification was published a year later. Jerry had
published something in one of the IEEE journals in 1976.

Then, I think it was 1981, Chuck manipulated things in the following
sense. Chuck's older boy, Danny, had a Bar Mitzvah. There was assigned
seating at the reception, and Chuck went out of his way to make sure
that I was seated next to Leo.

Leo and I talked for several hours about tree stuff. Somehow that led
to a manuscript, and that manuscript existed for quite a long time.
Some of it was medical stuff that I wrote and some of it was
mathematics. Regarding the latter, I wrote the initial draft; and Chuck
completely rewrote it.
Seven of the first eight chapters were from Leo. I remember vividly
Chuck saying that, ``with Leo the first 90 percent is easy and the last
10 percent is really hard. With me, if you can understand the notation,
it's all there.'' What happened is  that Leo took what Chuck wrote,
read it and he really didn't like~it.

\textbf{Rice:} But both pieces survived in the final book, right?

\textbf{Olshen:} Well, they did; but they survived in funny way. I've
told this story before in the pages of \textit{Statistical Science},
and I'll try to be brief. Basically what happened was at one point
Susan and I came up to Berkeley and were visiting Chuck for some reason
that I don't remember. We went to what used to be a very good open air
sandwich shop on Hearst, just below Euclid on the north side of the
street. The three of us ran into   Leo and Jerry. At that point, Leo
and Chuck hadn't spoken to each other for a long time, and the
manuscript lay dormant. Leo was always the gallant one and he said
``Why don't we get together after lunch at my office, and we'll hammer
this out?''

Susan said fine, and she had a book in her purse. She said, ``I'll go
to the library and read my book'' and I~said ``No you won't.'' I knew
that Leo had this gallant aspect to him, that Leo would never be harsh
in front of a woman. ``You're coming to our meeting.''

We came to Leo's office, the four of us; Jerry and I~were always
willing to compromise on almost any reasonable thing. Leo and Chuck
didn't get along all that well even though they were colleagues. I got
a chair and made sure that Susan sat on it between Leo and Chuck: Leo
and Jerry on one side; Chuck and me on the other.

I knew that if we were ever going to agree on anything that that was
the right environment, and we did. We came to some ground rules about
who was allowed to criticize whom, and that I would be the arbitrator.
I~would try to write things so that it read like a book, and make the
glossary and the table of contents and what have you. The book was
finished sometime in 1983 and was published in late 1983.

\textbf{Rice:} Yes, I still go back to it and read it for insights.
When I think I understand something, and then I realize that I don't, I
go back and read it again.

\textbf{Olshen:} We tried pretty hard. Of course now, it's somewhat
pass\'{e}. That was, of course, before boosting, though we
certainly realized that if you have a base rule for classification,
observations clearly marked for one class or the other aren't the hard
parts. The hard parts are observations near the boundary. The idea of
boosting made sense, but making science out of that is not a trivial matter.

I have the impression now there are lots of what I~consider pretty good
classifiers out there. There are neural nets done properly and support
vector machines, because Vapnik had this bully pulpit and wrote a book.
There's boosted CART. Then later Leo got into random forests. Those are
just some that come to mind.

I don't really think that the hard part of most classification problems
is whether you choose a support vector machine or boosted CART. I think
the hard part is knowing what features to include.

Knowing what features to include gets you the main digit in error rates
and risk. Whether it's support vector machine or boosted CART or
something else matters less; it's easy to fool any of them. But, if
you're any good at what you're doing you'll know, ``Gee, I don't think
I want to use a random forest for this because there are a lot of
features and most of them are noise; and I could fool it.'' Or, ``I
know the decision boundary is really smooth and a straight line so
vanilla CART doesn't make sense because the boundary doesn't have the
saw tooth.'' Well, you should know your subject matter well enough to
know that, and if you do you can usually be a pretty good guesser as to
what to use. But it matters whether you include this or its square or
the product of these two things or whatever.

\textbf{Rice:} Another hard thing about classification problems is not,
as you say, it's not whether you use support vector machines or random
forests, but how you actually construct the training set, where it
comes from, and what it's relation to the test set is. That's often
really quite nontrivial.

\textbf{Olshen:} Of course.

\textbf{Rice:} I think it's frequently glossed over.

\textbf{Olshen:} Well, the assumption of internal cross validation is
that the joint probability structure of the predictors and the outcome
are the same; and what you're testing is not. Does that make sense? In
a lot of applications, it doesn't. You see that all the time in
medicine. Just for an illustration: Suppose you have a truck and you go
to the county fair and you do mammograms. You could have some
classifier and it will be trained because you'll go to some medical
center and pull out 500 records of people who have breast cancer and
500 people who didn't. But in the county fair the prevalence/priors are
maybe one out of 500 or something like that. It's very different.
You're basically talking about different regions of the feature space,
different base rules, and the thing that worked for 500 versus 500 may
not work very well for one versus 250.

\textbf{Rice:} And the joint dependence structure of the covariates can
be different.

\textbf{Olshen:} Yes. In that part of the feature space, it might.
There are all kinds of things that can go wrong, and it's amazing that
in 2013 that one still needs to say such things out loud because these
are mistakes that are common today. It's not like, ``Oh, in olden days
people did things this fallacious way.'' Olden days may be 20 minutes ago.


\textbf{Rice:} Your interests changed. You went to Stanford; I think it
was in 1989. At some point in the School of Medicine, there you became
interested in genomics. That changed a lot of what you did. How did
that transition take place?

\textbf{Olshen:} Well, I'd always been interested in genetics. My first
wife and girlfriend who drew me to Berkeley in the first place wrote a
thesis about the genetics of mating latency in fruit flies. Basically,
the idea is that you had sites and people knew that the outcomes were
discrete. One of the big things on the table then was, ``How many genes
were involved? How many sites?'' These days you'd say, how many SNPs
were involved in producing a particular phenotype?

That's a deconvolution problem because the phenotype you see is the
product of some vector of genotypes plus some noise, you must
deconvolve the sum of things that matter and the noise.

I had a long standing interest in those problems. Then in the 1990s I
was very fortunate at Stanford, just as I had been in the Laboratory
for Mathematics and Statistics at UCSD, that I had very able
assistants. My assistant at Stanford, Bonnie Chung, said that I got a
phone call from Victor Dzau who was then the Chief of the Division of
Cardiology at Stanford and the Chair of the Department of Medicine.
Later he left Stanford.
He and others were starting a project that ultimately was called
SAPPHIRe, the Stanford Asia and Pacific Program on Hypertension and
Insulin Resistance. It involved people who I didn't know but should
have. David Botstein was one, and there were various others. Neil Risch
was somebody upon whom we could lean to help with calculations.

So for reasons that I don't know, Victor somehow got my name and I knew
who he was even though I doubt he knew much about me, and said, ``We're
going to be writing this grant Saturday morning.'' Well\ldots


I didn't take to being anyplace at eight o'clock Saturday morning. But
I finally got there at nine o'clock, having dragged myself out of bed,
because I realized that this was the big leagues; and even though
finding genes that predispose to hypertension is really tough, it
seemed like something I should get involved in. That was in the 1990s.
Since then things grew. The technology grew---one of my students worked
for a company, Affymetrix, that did a lot of SNP genotyping and
invented some of the technologies.

That technology was developed by a man in engineering and his daughter.
Part of my life has been in Electrical Engineering at Stanford. The man
is Fabian Pease. It involves embedding something in plastic and shining
laser light on what binds to it, the complimentarity of nucleic acids,
and the bending of laser light.

The bending of light leads to an inverse physical problem of making an
inference. I'm not going to go into details because there are other
places to read about it. But the point is that virtually all those
technologies, SNP technology, expression technology, and now protein
chips, in some sense they are all the same. Those are nifty problems.

They get harder the bigger the molecules you are embedding in the
plastic are. That's why the proteins are really tough. They tend to be
huge molecules, and they don't have very many binding sites.

I never got very much involved in gene expression, but I've certainly
been involved in the proteins and the actual SNPs themselves. Once
again, there's a triangle. There are SNPs; there is then gene
expression; and then the actual proteins that your body sees.

One thing has led to another, and a lot of problems have come up
related to that, one of them being immunology, very broadly defined.
That's how I got into this SAxCyB and protein arrays. The statistics of
it is not very foreign.

\textbf{Rice:} Another activity, of course, that has consumed your time
at Stanford and your interests is all your work on image compression
with Bob Gray and his colleagues. It's easy to see a path from CART to
that in broad brush. How did that begin?

\textbf{Olshen:} Well that started out because I was at Stanford on
sabbatical in 1987 and 1988. There was a graduate student in electrical
engineering named Phil Chou, who's now at Microsoft Research, a
brilliant person. Jerry Friedman, my CART colleague, was supposed to be
on his orals committee, and Jerry wasn't able to go to the exam. He
asked, ``Would you go?'' I was just a visitor, but it seemed of
interest and I went. Phil was clearly terrific. His thesis adviser was
Bob Gray, who was the master of compression. Bob's student Eve Riskin
saw that the pruning algorithm that's Chapter~10 of the CART book, that
came from the Technologies Services Corporation tech report, really
applied to image compression. Think of a binary tree and you could
think of bits telling you to go left or right, and you can think of the
average number of bits you need, and that's just the average depth of
the tree.

If you are building large trees and pruning them back, you'd be faced
with what amounts to the same problem in both cases. Anyway, when I
came back to Stanford in 1989, there was a phone call from Bob who was
looking for somebody with whom to collaborate, and he had problems in
image compression of various sorts. I was asked to help, and I did. I'm
not sorry I did; it's been an interesting chapter of my life.

We studied malignant masses in the mediastinum and in lungs by CT. We
studied flow through major blood vessels in the chest by MR. We studied
digital mammography, which turns out to be a really hard subject, and
also satellite images.

\textbf{Rice:} There's another thing you've been involved with at
Stanford that I know much less about, the Data Coordinating Center. You
haven't told me much about it in the past.

\textbf{Olshen:} Well, what I thought it was to be and the way it's
turned out aren't the same. My motivation was very simple. It used to
be that when anybody had his or her favorite algorithm for doing
classification of whatever, it was always, and I mean always, tried out
on the UC Irvine database.
I don't ever want to hear again about the UC Irvine database. I
thought, there's so much going on at Stanford. Why don't we just
organize something at Stanford and get Stanford data and use them for
standards in somebody's support vector machine or whatever? I decided
to organize something: the Data Coordinating Center. My hope sort of
panned out, and sort of did not. It still exists, but it's turned into
a boutique operation that does very fancy database things, mostly for
Stanford's Cancer Institute. Furthermore, HIPAA laws have intervened.
It's not a trivial matter to get data from somebody's experiment on
human beings to a statistician, or an engineer, or somebody who may
have something to say about, ``Yes, this person will get a malignant
disease,'' or, ``Yes, this person has hypertension,'' or whatever.

But I got that started before I knew the weight of HIPAA laws upon us.
My efforts were a reaction to my being sick after the 107th time that I
saw something from the Irvine database.
Some of the things I~was involved in at Stanford had to do with
nephrology, that is to say, with kidneys. I got involved with a group
in Phoenix; one of the NCI branches. NIDDK is there, and I worked with
a friend in his lab at Stanford. I knew that in the database at UC
Irvine is the Pima database. I knew that there are Pima Indians in
Arizona, because there's a reservation there. They have hardscrabble
biological cousins in northern Mexico who are skinny and not
hypertensive. The people in Arizona are insulin resistant, and they're
fat; and you can wonder why.

This seemed interesting because this suggested that there was some gene
by environment interaction going on, so that played into CART, into my
interest in that. It played into my history with nephrology, and I
realized, and maybe this is presumptuous of me, that probably many of
the people using the Pima Indian database in the UC Irvine collection
for testing their algorithms didn't know anything about hypertension,
or Pima Indians. That offends my aesthetic. Maybe it's because I'm so
poor computationally, but I've seen myself as a participant in people's
activities, but not more than that.

\textbf{Rice:} Let me probe a bit further into your role in
interdisciplinary studies. You've talked about several of them, and you
said one of the things you bring to them is humility; but actually, as
a statistician, you bring more. You're working with smart engineers, or
you're working with smart MDs, but you're bringing something as a statistician.

\textbf{Olshen:} I hope so.

\textbf{Rice:} You're bringing something to the table. I wonder if you
could articulate what you think that is.

\textbf{Olshen:} One answer might be an example. Something just came up
in the Workshop in Biostatistics, that I ran at Stanford for many
years, and for which I am now ably assisted by Chiara Sabatti, who does
most of the heavy lifting.

Imputation is a big deal in genetics these days. People make inferences
about the single nucleotide polymorphisms at sites for which they have
no data. They may actually sequence a half a million sites if they do a
lot, maybe many fewer if they are more specialized.

To impute they use something called haplotypes. My understanding of
what a haplotype is, is that there are long strings of DNA, and if I'm
at a given point and there are five points nearby and I know what those
are, then I must be part of such and such a cluster and, therefore, I
can read out fairly far. OK? What the genome is, then, is a bunch of
haplotypes strung together. I'm going to even forget about the
randomness of the fact that the partition of humanity is very coarse.
One can ask, ``What's the probability mechanism that generated these
things in the first place?''

After querying people in a large audience that included some people who
know genetics far better than I do, it seemed that because this
imputation is done with so called hidden Markov models, there needs to
be something that's at least approximately Markovian there. What is it?

I was able to get out of the discussion that what's Markovian are these
so-called haplotypes that get laid down. Well that means that the
marginal distribution of the individual sites is certainly not
Markovian. But what is it?

Well, people compute now the covariance function of sites. You can do
that, but then you have to ask yourself, is the covariance function you
compute consistent with that of a mixture of Markov processes? You
should be able to answer questions like that, because you should know
the probability mechanism that generated the data in the first place.

That's our job---to try to make those inferences. I~don't see those
kinds of questions being asked. You ask what I bring to the table,
maybe it's a sensitivity to things like what I've cited. That's an
example of something that's sort of statistical, sort of probabilistic.
One could think, ``What kind of tests would you do if you got data on
genotypes to figure out if something was a mixture of Markov processes
or not, and necessarily consistent with how haplotypes are said to be
generated?'' That's a question that it seems to me is worth asking. So
far as I can tell, it hasn't been asked.


\textbf{Rice:} I'm thinking about what you've just been saying about
this example and about numerous interactions with young people, both
statisticians, and nonstatisticians. I'm thinking particularly about
people who attend your biostat seminars, about graduate students and
post docs. What advice do you give them if they say, ``I'd like to be
doing this kind of thing, this interdisciplinary thing in the future.''
Do you tell them, ``Go out and learn about Markov processes?'' What do
you say?

\textbf{Olshen:} No. Well, first of all, hardly anybody ever asks. But
of those few who do, my only advice would be that anything you learn is
to the good. In particular, anything one can learn in mathematics is to
the good because it may come up in the future, and it certainly
sharpens the mind. Anything you can learn about the subject matter is
fine. But the most important thing you have to learn is you have to
learn how to learn, because, at least in my life, the things that I do
every day didn't exist as problems when I was a student. The world has
changed. I don't know if it has changed for the better, but it's
changed. One is constantly having to learn new things.

To summarize, the main things to learn are patience, learning how to
learn, learning how to be a student for the rest of your life. Because
if you go into some academic work, you are going to be a student for
the rest of your life, and not only that---I was speaking with Iain
Johnstone about this the other day because the question came up in
conversation---I think you have to enjoy the chase. The chase might
mean working on problem three in Chapter Seven.

It might mean the fact of trying to understand SNPs that are combined
with some environmental factors to predispose to insulin resistance or
hypertension. It might mean any one of a number of things. But if you
don't enjoy and get some charge out of just whatever the chase is, then
you are not going to be very happy; and you're probably not going to be
able to do much either, and there's a lot to do.

\begin{figure}

\includegraphics{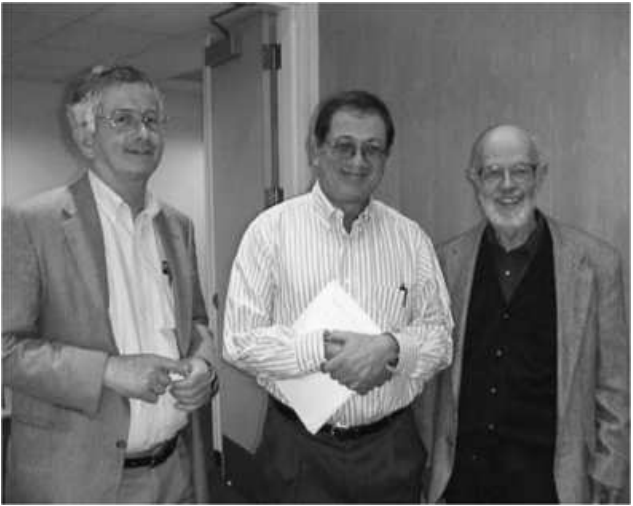}

\caption{Richard with Peter Bickel and Erich Lehmann at Berkeley in
2005. Peter is a long time friend and collaborator. The late Erich was
Richard's adviser his sophomore year at UC Berkeley in 1960--1961.}\label{fig9}
\end{figure}

\textbf{Rice:} You said you have to be a student. I think as you get
older it's hard to find the time to be a student.

\textbf{Olshen:} One has no choice.

\textbf{Rice:} You have to really want it, or else it's not going to happen.

\textbf{Olshen:} That's the only choice there is. One is a student. I
don't know what it would be like to be a super genius. But I can say
what's like to be me. If you just have maybe better than average but
not such spectacular gifts, then you just have to be willing to plug
away and to be patient and cross your fingers, and hope for the best.
But one of the things, also, that I think, because this has come up in
conversations far removed from this discussion lately is this: it's
really nice when people come along afterwards and they come up with a
simple proof of something. You think, ``That's great.'' But the first
person that got there didn't know, didn't know what the answer was.
Maybe yes, maybe no, maybe this, maybe that. To me, that's the hard
part and the fun part of every subject. In that respect, there is no
disconnect between medicine and mathematics. They are just hard things
to do. They're things one doesn't understand and one crosses ones
fingers and hopes that one will learn to explain some phenomenon. I can
say in my case that I've certainly been disappointed many times. That
maybe it's just because I've made unfortunate choices.

But I think the people who are most successful have been successful at
least in part because they've been wise about how to spend their time.
Everybody's only got so much time. There are a few super geniuses, but
there are not enough to populate all the universities.

But some people are clearly better than others at picking things to
work on. Afterward it's easy to say, ``If I had thought of that\ldots'' Well, the point is that you didn't.

Well it's just like you and Bernard's finding the eigenfunctions and my
gait stuff. After the fact, I see   that's a kind of obvious thing
to do. Not that I know how to form confidence intervals for those
predictions very well. A lot of things are easier in hindsight than
they were in foresight.

\textbf{Rice:} Yes. Foresight's limited. I was thinking about yours. I
was trying to put myself in your position when you were working on the
fluctuations of periodograms. Then in light of things we've just been
talking about, if you try to look ahead from your point of view then,
what things would most surprise you about statistics? There've been
lots of changes. Are there particular things which you just wouldn't
have envisioned, which have surprised you especially?


\textbf{Olshen:} I think that modern computing has changed the world.
It will never be the same, and it shouldn't be. I think that it's not
settled yet. Because there is one view of this world that says, ``Well,
I don't know if this is a good model for $x$, $y$ or $z$; so I'll simulate.''
You'll simulate five cases and they'll all come out heads! I can toss a
coin five times and it'll come out all heads, and I'll think that's a
two headed coin; but maybe that just happens that it came out heads
five times. Then on the other hand if you're just curmudgeonly and say,
``Well I~won't believe a word unless I can prove some theorem about
it,'' you almost never can. What is the right way to be? I have no
idea. Maybe 100 years from now, if the world doesn't blow itself up or
poison itself, then maybe people will figure that out better.

I'm 70 years old. Actuarial chances are that I'm not going to live that
much longer, and my health isn't so terrific. We just get just this
little slice of time. I'm not a hyper-religious person, but I do try to
read the Torah portion in the Old Testament every week.

The Hebrew is really beautiful. I can't translate a lot of it, but what
I can translate is really good. Not only that, but in books that one
reads, one realizes that hundreds and thousands of years ago there were
some really smart people who wrote great stuff. As Bradley Efron
reminds me, if Mozart hadn't lived, it isn't that somebody else would
have written ``Don Giovanni.'' We wouldn't have ``Don Giovanni.''

But whatever we've done, and nobody in science does that much because
you know it'll all be rediscovered somehow, in some fashion anyway.
What we know from the past is just a distillation of what happened. Who
knows if what's distilled and thought to be so nice now was in its day
thought to be so nice!

I've had occasion recently to be interested in the distribution of the
sum of independent uniform random variables. It just came up as a
matter of so-called meta analysis and all this. It's not a trivial
matter, because you can think of picking a point on a hypercube and a
plane sliding through the hypercube. But hypercubes have corners, and
they screw up distributions. Well, so I've learned that in 1920s two
very smart people, one named J. O. Irwin and the other Philip Hall, who
went on to become a famous mathematician, figured out how to do that.
They published in \textit{Biometrika}.

\textbf{Rice:} Figured out how to do what?

\textbf{Olshen:} How to compute the distribution of the sum of IID
uniforms. That sounds like a simple exercise, but just try to do it.
It's not so simple. You can invert a Fourier transform if you're good
at inverting Fourier transforms, but that involves complex integrals.
It turns out that the essential computation for that, and this is a
footnote that Karl Pearson put in \textit{Biometrika}, the essential
computation that enables you to compute the distribution of the sum of
IID uniforms was done by Euler, who apparently didn't know anything
about applications and could not have cared less. Good for him, but was
that worth anything in those days?

How did I get interested in that? I got interested in it because it had
to do with combining independent tests into one test of significance.
If you think that the null hypotheses are true, then you've got a
uniform draw on the unit interval. You've got, collectively, a point on
the unit cube. Fisher's minus twice summation thing results in a
hyperbolic neighborhood of zero. What if you wanted a linear
neighborhood of zero? This is something my son Adam got me into.

Well, the question is easy to state, but the answers aren't always easy
to come by. I think that what is even the right thing to do in given
applications is far from obvious.

\textbf{Rice:} You have been quite a valuable mentor to young people.
Is there anything you can say about that
process?

\textbf{Olshen:} There are few guidelines. It says in Torah that there
are two classes of people in the world of whom you must never be
jealous, your children and your students. That's one set of guidelines.

Another thing: some cultures have a severe, if implicit, concern about
respect for elders; whereas Jewish culture has in it a healthy
skepticism of the wisdom of elders. Now that I am old, I wouldn't mind
a little more respect; but I think that it can be overdone because the
future is for young people.

My attitude is that no future was built on the backs of 70-year olds.
The future is in young people. If you think that the young people are
what will become us (and we won't be here to see what they do) then you
would like for them to look back on you perhaps favorably to the extent
that what you instilled in them was something worthwhile.

\textbf{Rice:} You've been in academic institutions representing
statistics in one way or another, depending on the institution.
Academic structures and education are changing, the roles of statistics
can be different in different universities, depending on the
environment, and those environments are changing.

\textbf{Olshen:} I think statistics is in a really difficult place,
because it has to justify itself as having something of its own, on the
one hand, and being a servant of other fields on the other. You and I
have talked about that. I think   that's a scenario that is hard for
university administrators to understand.

\textbf{Rice:} Well, it's a strength and simultaneously a weakness.

\textbf{Olshen:} That's true. It's a perpetual problem, and I~don't
think it's going to go away. However, there are other people trying to
eat our lunch. Computer science is, for example. To me it is about data
structures and related subjects. These are fields about which
statisticians could do well to know more. However, to too great an
extent, computer science is rediscovering the wheel. I think that in
classification, for example, or machine learning, there is much too
much encroachment by computer scientists.

\textbf{Rice:} What are your plans for the future? What are you looking
forward to doing?

\textbf{Olshen:} Don't know. I think about that, but I have no idea. I
mean, I realize that one useful purpose I can serve is to be a
babysitter for grandchildren. That's important. That's clearly a task
that I am deemed able to do.

\textbf{Rice:} Congratulations.

\textbf{Olshen:} Beyond that? I don't know, more of the same. I'm
trying to get some papers done now. I can't run as fast as I used to. I
used to be sharper than I am now. All I've ever had is just the ability
to react to situations that weren't always of my choosing and weren't
always enviable either. My health is pretty poor.

I'm trying to write a monograph on the successive normalization of
rectangular arrays of numbers, and I~see there's lots to do, and I
don't know if I'll get to that. But I hope to.

\textbf{Rice:} Well maybe it gets back to Yogi Berra, right? It's hard
to predict what's going to interest you in the future. Would you have
predicted five years ago that you'd be interested in normalizing
rectangular arrays? Probably not.

\textbf{Olshen:} No. That came up as a challenging mathematical
problem. But I see that it has practical consequences. It's like making
inferences about vectorial data, whether you look at covariances or
correlations, you learn different things from each one; and that's
inescapable. I'm also trying to rewrite something for some referees now
that has to do with defining insulin resistance rigorously and finding
if there are SNPs and candidate genes that predispose to it. I just
finished something with my son, Adam, on ribosomal profiling.

There's another project that has to do with HIV. HIV used to be an
acute disease and you'd get it and you were dead quickly. Drugs now
really prolong life, but they are pretty potent stuff. They're pretty
bad, and you have to worry. If somebody is going to be alive for 10, or
15, or 20 or 30 years, you'd better worry about whether the potion you
are giving is going to cause heart disease, or kidney disease or
something else. There are ways of trying to make those inferences.

\textbf{Rice:} We're very fortunate to be in a profession with so many
opportunities, aren't we?

\textbf{Olshen:} Yes, it's a pretty good deal. I remember in San Diego
at the Rosenblatt's house many years ago, the late Errett Bishop asked,
``What would you do if you could do anything? Would you work in
algebraic geometry, do this or do that\ldots?''\newpage

\textbf{Rice:} Or, constructive mathematics. Of course!

\textbf{Olshen:} I said, ``Errett, I would do exactly what I'm doing. I
would just be better at it because I'd be smarter.''

\textbf{Rice:} He must have been very disappointed by that answer.

\textbf{Olshen:} Disappointed? He didn't believe me! But that's what I
think. I told him, I said, ``I'd do exactly what I'm doing. I'd just be
better at it.'' He was very upset; he didn't like that at all. But I
thought that was an honest reply. I think that a lot of people who have
jobs as statisticians of some form or other deep down believe that.
That's how they conduct their lives. Unfortunately, it's going to be an
ongoing necessity to justify ones existence as a statistician; but it
is an honorable way to conduct your life.




\section*{Acknowledgments}

This interview took place in January 2013, in Berkeley, California. We
are very much indebted to
Frank Samaniego for his encouragement and assistance, and Bonnie Chung
for technical assistance.

\begin{figure}[t]

\includegraphics{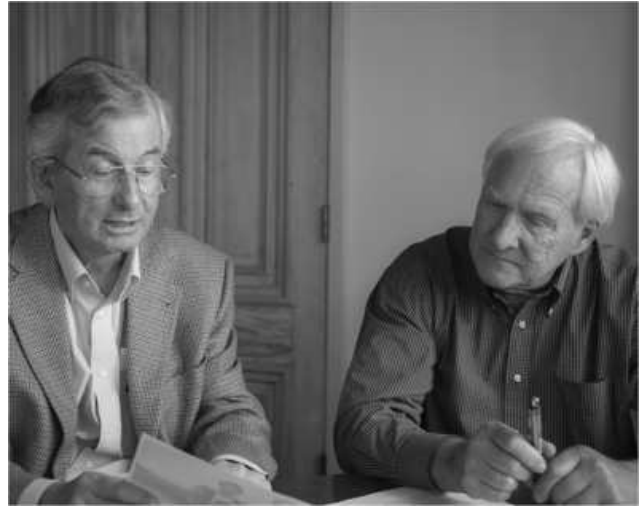}

\caption{Richard Olshen and John Rice in the Fall of 2013, Berkeley, CA.}\label{fig10}
\end{figure}


\end{document}